\documentclass[11pt]{article}
\usepackage{moriond,epsfig}

\def\Journal#1#2#3#4{{#1} {\bf #2}, #3 (#4)}

\def\NPB{{\em Nucl. Phys.} B}
\def\PLB{{\em Phys. Lett.}  B}
\def\PRL{\em Phys. Rev. Lett.}
\def\PRD{{\em Phys. Rev.} D}
\def\ZPC{{\em Z. Phys.} C}

\def\be{\begin{equation}}
\def\ee{\end{equation}}
\def\bea{\begin{eqnarray}}
\def\eea{\end{eqnarray}}

\begin{document}
\vspace*{4cm}
\title{KLOE RESULTS ON $f_0$(980), $a_0$(980) SCALARS AND $\eta$ DECAYS}

\author{THE KLOE COLLABORATION\\ $\ $ \\
{
  F.~AMBROSINO, A.~ANTONELLI, M.~ANTONELLI, C.~BACCI, P.~BELTRAME,
  G.~BENCIVENNI, S.~BERTOLUCCI, C.~BINI, C.~BLOISE, V.~BOCCI,
  F.~BOSSI, D.~BOWRING, P.~BRANCHINI, R.~CALOI, P.~CAMPANA, 
  G.~CAPON, T.~CAPUSSELA, F.~CERADINI, S.~CHI, G.~CHIEFARI,
  P.~CIAMBRONE, S.~CONETTI, E.~DE~LUCIA, A.~DE~SANTIS, P.~DE~SIMONE,
  G.~DE~ZORZI, S.~DELL'AGNELLO, A.~DENIG, A.~DI~DOMENICO, 
  C.~DI~DONATO, S.~DI~FALCO, B.~DI~MICCO, A.~DORIA, M.~DREUCCI, 
  G.~FELICI, A.~FERRARI, M.~L.~FERRER, G.~FINOCCHIARO, C.~FORTI, 
  P.~FRANZINI, C.~GATTI, P.~GAUZZI, S.~GIOVANNELLA, E.~GORINI, 
  E.~GRAZIANI, M.~INCAGLI, W.~KLUGE, V.~KULIKOV, F.~LACAVA, 
  G.~LANFRANCHI, J.~LEE-FRANZINI, D.~LEONE, M.~MARTINI, 
  P.~MASSAROTTI, W.~MEI, S.~MEOLA, S.~MISCETTI, M.~MOULSON, 
  S.~M\"ULLER, F.~MURTAS, M.~NAPOLITANO, F.~NGUYEN, M.~PALUTAN, 
  E.~PASQUALUCCI, A.~PASSERI, V.~PATERA, F.~PERFETTO, L.~PONTECORVO, 
  M.~PRIMAVERA, P.~SANTANGELO, E.~SANTOVETTI, G.~SARACINO, 
  B.~SCIASCIA, A.~SCIUBBA, F.~SCURI, I.~SFILIGOI, T.~SPADARO, 
  M.~TESTA, L.~TORTORA, P.~VALENTE, B.~VALERIANI, G.~VENANZONI, 
  S.~VENEZIANO, A.~VENTURA, R.VERSACI, G.~XU}\\ $\ $ \\
  presented by S. GIOVANNELLA}

\address{Laboratori Nazionali di Frascati dell'INFN, via Enrico Fermi 40, 
  00044 Roma, Italy}

\maketitle\abstracts{
  The KLOE experiment running at the $\phi$-factory DA$\Phi$NE has 
  collected $\rm \sim 450\ pb^{-1}$ in the 2001--2002 data taking.
  We report preliminary results on light meson spectroscopy based
  on this data sample; particles are all produced through $\phi$ 
  radiative decays.
  The nature of $f_0$(980) and $a_0$(980) is investigated by studying
  the shape of the resulting mass spectra, which is sensitive to their 
  structure.
%
  A detailed study of the $\eta\to\pi\pi\pi$ dynamics through
  a Dalitz plot analysis gives the possibility to extract information
  on the quark mass difference.
  Finally, the branching ratio for the $\eta\to\pi^0\gamma\gamma$ decay
  is compared with previous measurements and with the expectations from
  Chiral Perturbation Theory.
}

\section{Introduction}

The KLOE experiment \cite{KLOE} operates at DA${\Phi}$NE, \cite{DAFNE} 
the Frascati $e^+e^-$ collider, whose center of mass energy is equal
to the $\phi$ mass. Data collected in 2001-2002, corresponding to 
$\sim 450$ pb$^{-1}$, are used to study light scalar and pseudoscalar 
mesons produced through $\phi$ radiative decays. 


\section{Light Scalar Mesons: $f_0$(980) and $a_0$(980)}

A complete study of the radiative decay of the $\phi$ to the scalar 
mesons $f_0(980)$ and $a_0(980)$ is in progress, involving the decays 
$f_0\to\pi^+\pi^-/\pi^0\pi^0$ and $a_0\to\eta\pi^0$, with 
$\eta\to\gamma\gamma$ and $\eta\to\pi^+\pi^-\pi^0$.
Since the mass spectra are sensitive to the nature of such mesons
\cite{AchIvanch} which are still puzzling,~\cite{qqbar,qqqq,kkbar} the 
data are compared with two different theoretical models. In the first one
the scalar amplitude is described by the kaon-loop model~\cite{SigmaKloop}
while in the second one a  point-like approach
is followed. In both cases, the interference with  
the background with the same final state is taken into
account in the fit procedure.

\begin{figure}
\begin{center}
\begin{tabular}{cc}
\psfig{figure=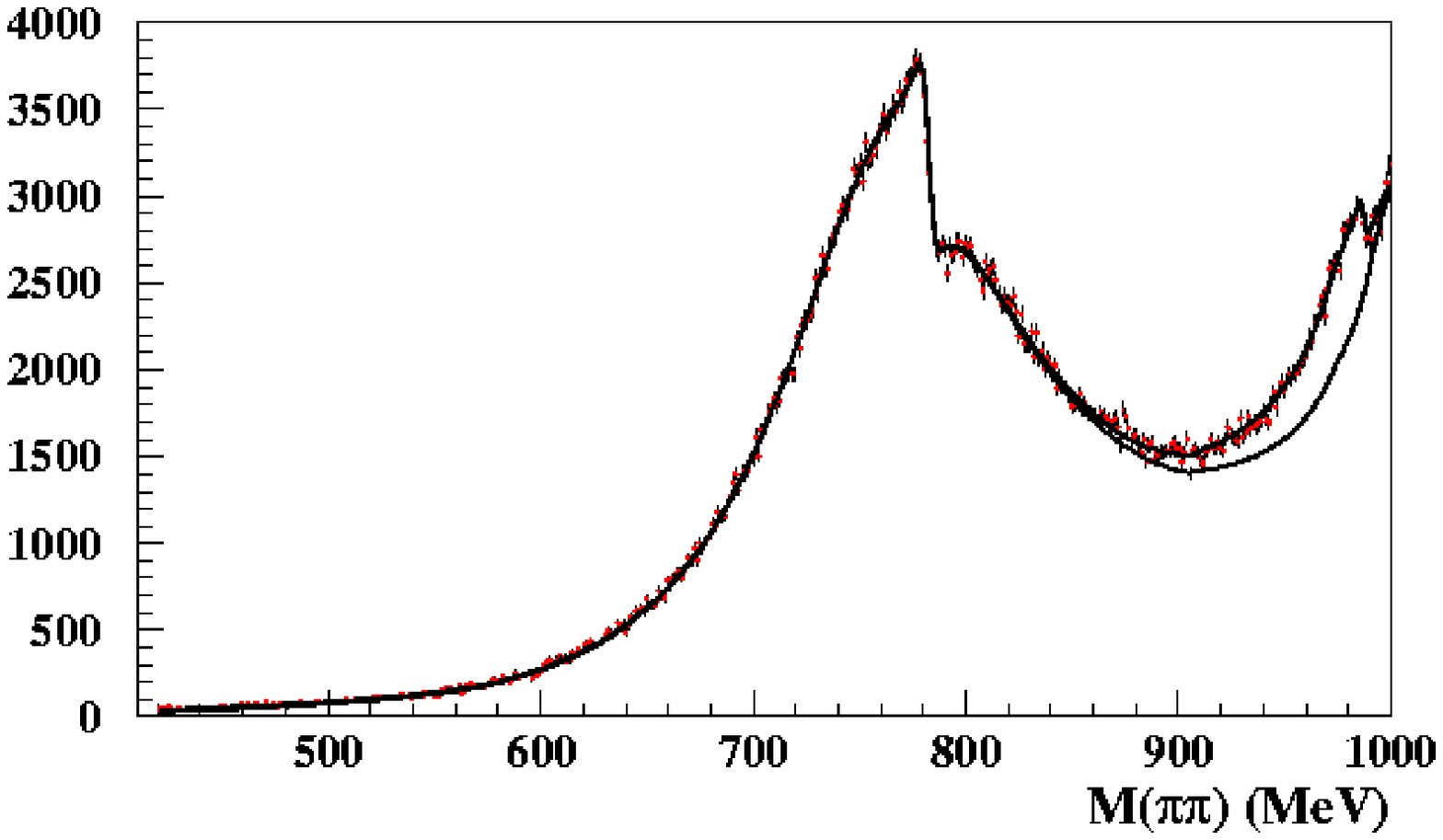,width=0.5\textwidth,height=6cm} &
\psfig{figure=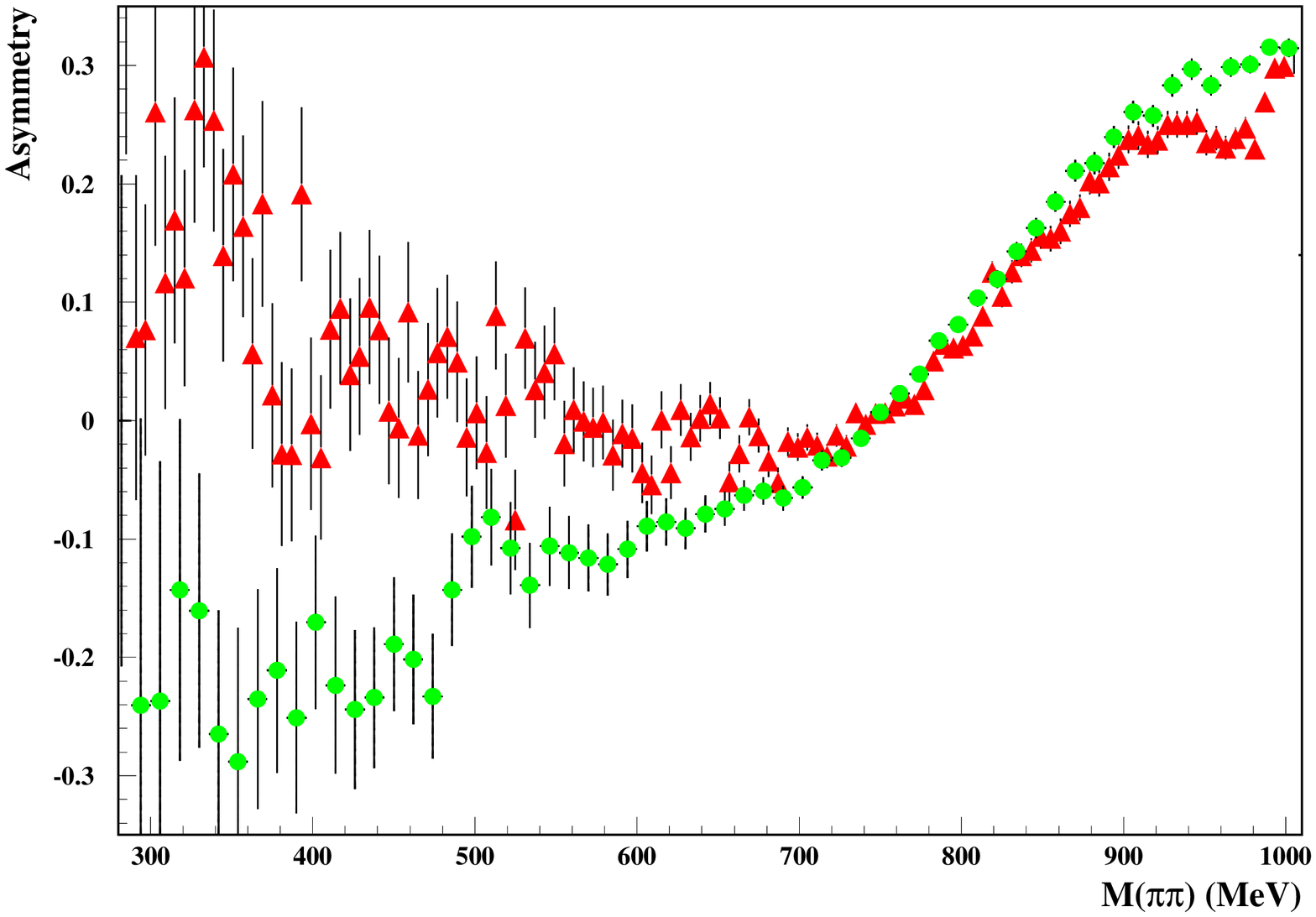,width=0.45\textwidth,height=6cm} \\
\end{tabular}
\end{center}

\vspace{-0.2cm}
\caption{Left: two pion invariant mass for $\pi^+\pi^-\gamma$ events.
  The upper and lower curves are the result of the fit and the
  contribution due to FSR+ISR respectively.
  Right: forward-backward asymmetry as a function of $M_{\pi\pi}$.
  Experimental data are reported as dark triangles while light dots 
  represent the Monte Carlo expectations for FSR and ISR only.\hfill
  \label{Fig:ppg}}
\end{figure}

For the $\pi^+\pi^-\gamma$ final state there is a huge irreducible 
background of $e^+ e^- \to \pi^+ \pi^-$ with an additional photon
due to initial state (ISR) or final state radiation (FSR). However, 
requiring two tracks and a large angle photon a clean signal appears
in the $M_{\pi\pi}$ region above 850 MeV (see Fig.~\ref{Fig:ppg}.left).
Moreover, a forward-backward asymmetry 
$A=\frac{N^+(\theta > 90^\circ)-N^+(\theta < 90^\circ)}
{N^+(\theta > 90^\circ)+N^+(\theta < 90^\circ)}$
is expected due to the interference of FSR and ISR.~\cite{Asymm} 
In Fig.~\ref{Fig:ppg}.right we show this asymmetry as a function 
of $M_{\pi\pi}$, both for data and for theoretical predictions with 
ISR and FSR only. A clear discrepancy is observed in the $f_0$ region 
and in the mass range below 700 MeV, thus adding a further evidence
on the need of a scalar meson in the theoretical description.

In the case of the $f_0\to\pi^0\pi^0$ decay we instead deal with a 
non-resonant  background with the same $\pi^0\pi^0\gamma$ signature,
produced  through $\omega\pi^0/\rho\pi^0$ intermediate states.
The intensity of this background is twice the signal. In order
to consider its interference with the scalar term we fit the Dalitz 
plot distribution. A smaller background contamination dominated by 
$\phi\to\eta\gamma$, with $\eta\to\pi^0\pi^0\pi^0$ and two lost
or merged photons, is estimated by Monte Carlo and subtracted from 
the Dalitz plot. When using the kaon-loop model we cannot describe data 
without introducing a scalar term due to a $\sigma(600)$ meson.

For the fully neutral search of $\phi \to  a_0 \gamma$,
the background with the same $\eta\pi^0\gamma$ final state is 
small and simplifies the fit procedure.
On the other hand, having a yield ten times smaller than the
$f_0\to\pi^0\pi^0$, it is contaminated by a large non-interfering 
background with a five photon signature.
The $a_0$ decay chain with $\eta\to\pi^+\pi^-\pi^0$ has instead
a rate three times smaller than the neutral channel, but it is 
completely background free.
%
A combined fit of the two channels is in progress to extract 
the $a_0$ parameters.

\section{Dynamics of $\eta\to\pi\pi\pi$}

The amplitude of $\eta\to\pi\pi\pi$ is related to the d-u
quark mass difference; a precise study of 
this decay can lead to a very accurate measurement of 
$Q^2=(m_s^2-\hat{m}^2)/(m_d^2-m_u^2)$.
Using the 17 millions $\eta$ mesons produced in 2001/2002,
the dynamics of both $\pi^+\pi^-\pi^0$ and $\pi^0\pi^0\pi^0$ 
final states has been studied through a Dalitz plot analysis. 
The $\eta$ mesons are clearly tagged by detecting the 
monochromatic recoil photon of the $\phi\to\eta\gamma$ decay 
($E_{\rm recoil}=363$ MeV); the background is at the level of 
few per mill.

Concerning the $\pi^+\pi^-\pi^0$ final state, the conventional
Dalitz variables are $X\propto T_+ - T_-$ and $Y\propto T_0$, 
where $T$ is the kinetic energy of the pions. The measured 
distribution is parametrized as:
$|A(X,Y)|^2 = 1 + aY + bY^2 + cX + dX^2 + eXY + fY^3$.
As expected from $C$ parity conservation, the odd powers of $X$ 
are consistent with zero (see Tab.~\ref{Tab:FitDalitz}).
Using our fitted parameters, the value of $Q$ can be extracted.
For example, in Ref.~9 the value $Q=22.8\pm 0.4$ is obtained, the 
error being dominated by the $\eta\to\pi^+\pi^-\pi^0$ width.
This value is in agreement with Chiral Perturbation Theory
($\chi_{PT}$) predictions \cite{Qtheory} and with other evaluations
based on $\eta$ decays, \cite{Qeta1,Qeta2} which have larger errors.

\begin{table}[t]
\caption{Fitted parameters of the $\eta\to\pi^+\pi^-\pi^0$ Dalitz 
  plot.\label{Tab:FitDalitz}}
\vspace{0.4cm}
\begin{center}
\begin{tabular}{|c|c|c|}\hline
  $\rm N_{dof}$ & Prob($\chi^2$) (\%) & 
  \begin{tabular}{c} 
    $a$ \phantom{00000000000000000} $b$ \phantom{00000000000000000} $c$ \\
    $d$ \phantom{00000000000000000} $e$ \phantom{00000000000000000} $f$ \\ \end{tabular}\\ \hline
  147  &  60   & 
  \begin{tabular}{ccc} $-1.072\pm0.006^{+0.005^{\phantom{0}}}_{-0.007_{\phantom{0}}}$ &
                 $0.117\pm0.006^{+0.004}_{-0.006}$ & 
                 $0.0001\pm0.0029^{+0.0003}_{-0.0021}$ \\ 
                 $\phantom{-}0.047\pm0.006^{+0.004^{\phantom{0}}}_{-0.005_{\phantom{0}}}$ &
                 $0.006\pm0.008^{+0.013}_{-0.000}$ &
                 $0.13\phantom{00}\pm0.01\phantom{00}^{+0.02\phantom{00}}_{-0.01}$ \\ \end{tabular}\\ \hline 
  149  &  63   & 
  \begin{tabular}{ccc} $-1.072\pm0.005^{+0.005^{\phantom{0}}}_{-0.008_{\phantom{0}}}$ & 
                 $0.117\pm0.006^{+0.004}_{-0.006}$ & --- \\ 
                 $\phantom{-}0.047\pm0.006^{+0.004^{\phantom{0}}}_{-0.005_{\phantom{0}}}$ & --- & 
                 $0.13\phantom{00}\pm0.01\phantom{00}^{+0.02\phantom{00}}_{-0.01}$ \\ \end{tabular}\\ \hline 
\end{tabular}
\end{center}
\end{table}

For the $\eta\to\pi^0\pi^0\pi^0$ decay the Dalitz plot density is
described by a single parameter $\alpha$: $|A|^2\propto 1+2\,\alpha\, z$, 
where $z$ is related to the three pion energies in the $\eta$ 
rest frame. 
Photons are paired to $\pi^0$'s after kinematically constraining
the total 4-momentum to $M_\phi$, thus improving the energy resolution.
By fitting a sample with high purity on pairing (98.5\%), corresponding 
to an analysis efficiency of 4.5\%, we get:
\begin{equation}
{\rm \alpha = -0.013 \pm 0.005_{stat} \pm 0.004_{\,syst} }\,.
\label{Eq:Alpha}
\end{equation}

\section{The $\eta\to\pi^0\gamma\gamma$ Decay}

The $\eta\to\pi^0\gamma\gamma$ decay is an important test of $\chi_{PT}$
because of its sensitivity to $p^6$ on both the branching ratio (BR) and 
the $M_{\gamma\gamma}$ spectrum.~\cite{ChPT1,ChPT2}
The present experimental situation is not completely clear: the most 
accurate determination of the BR \cite{GAMS} is far from theoretical 
predictions while a more recent measurement, \cite{CrystalBall} with a 
larger relative error, gives a significantly lower value.
Moreover, all previous searches were done at hadron machines, using 
mainly $\pi^- p\to \eta\, n$. The value of the BR has decreased by three 
orders of magnitude in the last 40 years, due to the improved separation 
of the $\eta\to\pi^0\pi^0\pi^0$ background.

\begin{figure}
\begin{center}
\psfig{figure=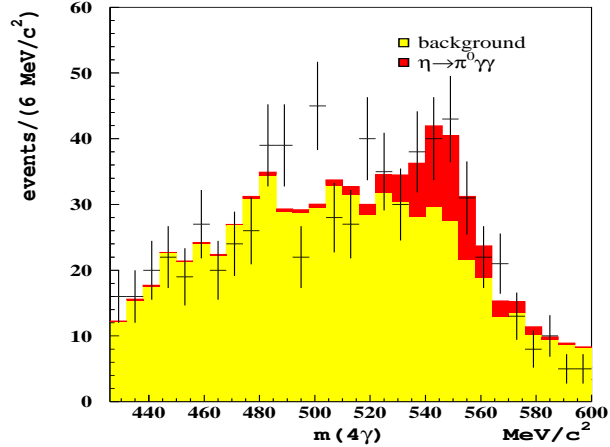,width=8cm,height=6cm}
\end{center}
\caption{Four photon invariant mass for $\eta\to\pi^0\gamma\gamma$ events.
  Data (crosses) are fitted with the signal and background contributions 
  evaluated from MC (solid histograms).\hfill
\label{Fig:Minv_pi0gg}}
\end{figure}

KLOE searches for this decay in a much cleaner environment, with 
different background topologies and experimental systematics. 
The two orders of magnitude higher background with the same five photon 
final state ($e^+e^-\to\omega\pi^0\to\pi^0\gamma\pi^0$, 
$\phi\to f_0\gamma\to\pi^0\pi^0\gamma$, 
$\phi\to a_0\gamma\to\eta\pi^0\gamma$ with $\eta\to\gamma\gamma$)
is reduced by vetoing the additional $\omega/\pi^0/\eta$ particles in
the event.
The remaining background is $\eta\to\gamma\gamma$ with additional
clusters from shower fragmentation or machine background and
$\eta\to\pi^0\pi^0\pi^0$ with merged/lost photons. 
We reject them with energy momentum conservation and a likelihood 
technique to identify merged clusters.
The preliminary results obtained fitting the $\eta$ invariant mass 
spectrum (Fig.~\ref{Fig:Minv_pi0gg}) gives a BR in agreement with 
${\cal O}(p^6)$ $\chi_{PT}$ calculations, with a central value which 
is three times smaller than the previous measurement:
\begin{equation}
{\rm BR(\eta\to\pi^0\gamma\gamma) = 
  ( 8.4 \pm 2.7_{stat} \pm 1.4_{\,syst}) \times 10^{-5} }\, .
\label{Eq:BRpi0gg}
\end{equation}

\section*{References}

\end{document}